\documentclass[prb,showpacs,preprintnumbers,amsmath,amssymb,superscriptaddress]{revtex4}
\usepackage{epsfig}
\usepackage{graphicx}
\usepackage{dcolumn}
\usepackage{bm}

\begin{document}

\title{Magnetic state and electronic structure of plutonium from ``first principles'' calculations}

\author{V.I.~Anisimov, A.O.~Shorikov}
\affiliation{Institute of Metal Physics, Russian Academy of Sciences-Ural Division,
620219 Yekaterinburg GSP-170, Russia}
\author{J. Kune\v{s}}
\affiliation{Institute for Physics, University of Augsburg, Augsburg 
D-86135, Germany}
\affiliation{Institute of Physics,
Academy of Sciences of the Czech Republic, Cukrovarnick\'a 10,
162 53 Praha 6, Czech Republic}

\date{\today}

\begin{abstract}
The goal to describe plutonium phases from ``first principles'' calculation methods is complicated by the problem of 5f-electrons localization. While for early actinides (Th, U,Np) standard DFT  (Density Functional Theory) description with itinerant 5f-electrons works well for late actinides (Am, Cm) DFT calculations with 
completely localized (pseudocore) 5f-electrons give satisfactory results. However plutonium presents a border case of partial localization and both limits (itinerant and completely localized) are not valid. We present a review of the methods used to solve this problem and discuss what could be the reasons for their successes and failures.    
\end{abstract}

\pacs{71.27.+a, 71.70.-d, 71.20.-b}

\maketitle

\section{Introduction}
The Density Functional Theory (DFT)\cite{DFT} in Local Density Approximation (LDA) (or in more elaborated Generalized Gradient Approximation (GGA)) has been very successful in describing equilibrium volumes and stable crystal structures for pure elements and its compounds. While there are sometimes difficulties in reproducing spectral properties, such as energy gap values for semiconductors and insulators, which are defined by excited electronic states, ground state energy as a function of the crystal volume and atomic positions is reproduced usually quite well. For rare earth elements problems appears in DFT calculations when one treats 4f-electrons on equal footing with all others. However in this case a good approximation is to consider 4f states as completely localized in the same way as core orbitals (so called pseudocore approximation). 

\begin{figure}[!t]

\begin{center}
\epsfxsize=14cm
\epsfbox{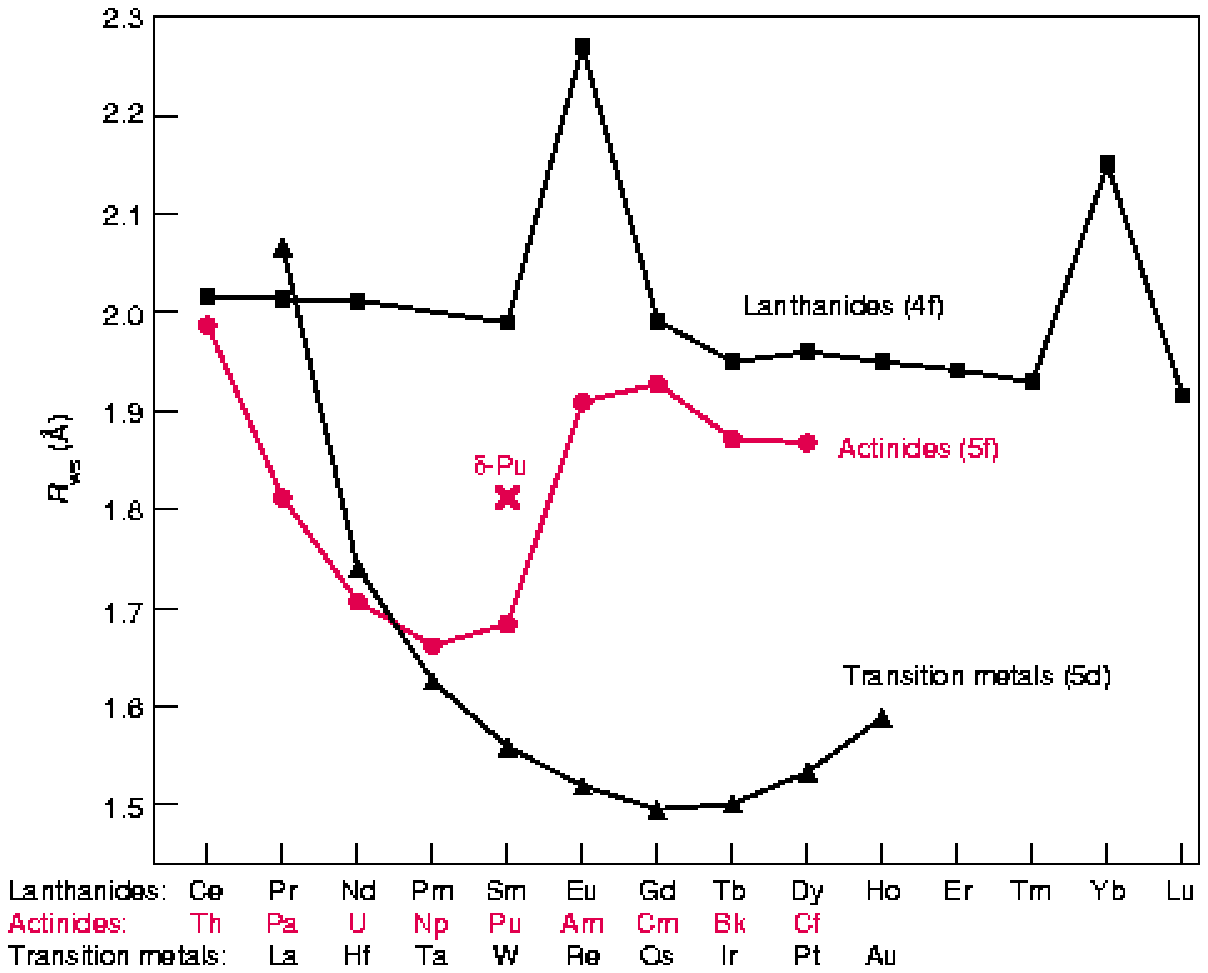}
\end{center}
\caption{Experimental values of equilibrium volume for 5d, 4f and 5f elements as a function of number of electrons in partially filled shell \cite{LosAlamos}.}
\label{fig1}
\end{figure}

Actinide elements give a challenge to DFT because 5f-electrons are not obviously localized as 4f electrons in rare earth but also can not be always considered as itinerant. On the Fig.\ref{fig1}  experimental values of equilibrium volume for 5d, 4f and 5f elements are presented. One can see that 5d elements show decreasing of the volume as number of 5d electrons increases till 5d shell becomes half filled. Further filling of 5d shell results in increasing of the volume. That corresponds to the strong participation of 5d electrons in the chemical bonding of transition metals evidencing their fully itinerant behavior suitable for treatment by DFT. In contrast to that lanthanides show very weak dependence of the volume on 4f shell filling (with the exception of Eu and Yb case where valency state is +2 instead of the common for lanthanides +3). That fact can be explained by fully localized nature of 4f electrons and hence absence of their contribution to chemical bonding. In actinides series beginning of the curve follows transition metals pattern: decreasing of the volume with increased filling of 5f-shell from Th till Np. However with further increasing of the number of 5f-electrons there is a sizable jump of the volume value for delta-Pu and Am and for late actinides Cm,Bk and Cf one observes weak dependence of the volume on 5f-shell filling resembling lanthanides with localized 4f electrons. This curve can be understood if one assumes itinerant nature of 5f-electrons from Th to Np, localized 5f states from Am to Cf and intermediate situation that can be described as ``partial 5f-electrons localization'' for Pu\cite{Freeman}. The problem is complicated by the fact that Pu itself exists in 6 various allotropies  with volume values differences reaching 20\% (e.g. between alpha and delta phases). That can be interpreted as more localized 5f-electrons in delta phase than in alpha phase.

\section{Standard density functional calculations}

\begin{figure}[!t]

\begin{center}
\epsfxsize=14cm
\epsfbox{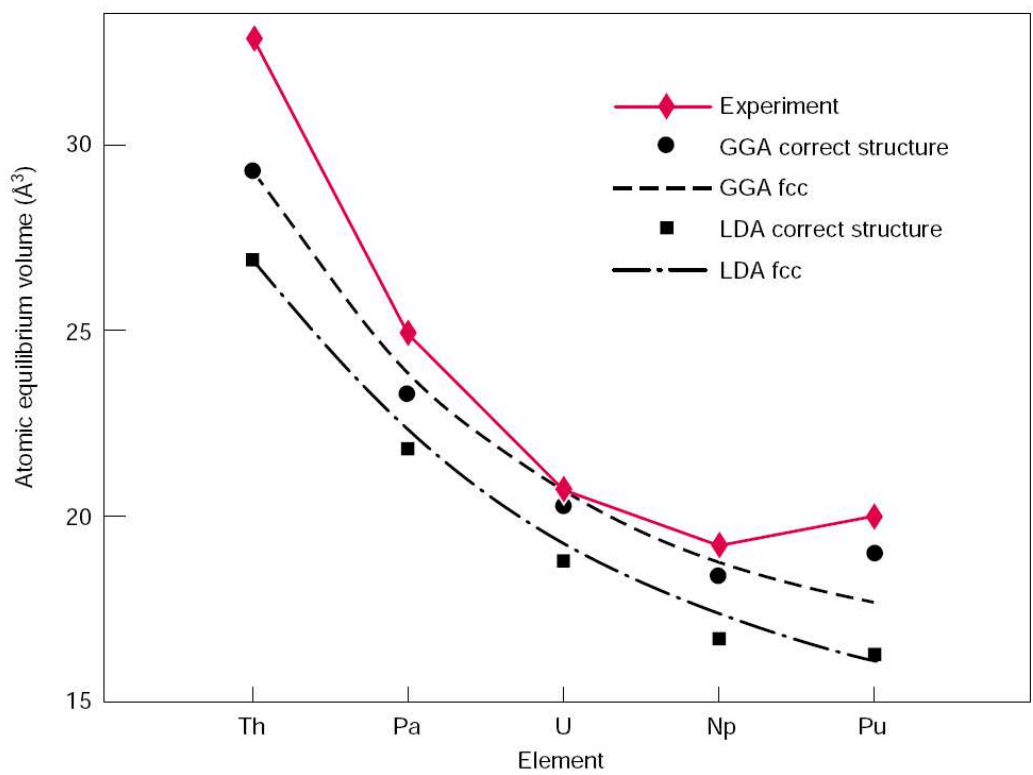}
\end{center}
\caption{Theoretical values of equilibrium volume for 5f elements 
from Density Functional Theory calculations\cite{LosAlamos}.}
\label{fig2}
\end{figure}

Standard DFT calculations for experimentally observed paramagnetic ground state gave good results for equilibrium volumes of early actinides  from Th till Np (Fig.\ref{fig2}). However already for alpha phase there is an underestimation of theoretical volume comparing with experimental one and for delta phase the disagreement is rather large (more than 20 \%). These results agree well with the above analysis of the experimental data for equilibrium volume of actinides. In DFT all electrons are considered to be itinerant and participating in chemical bonding. That is a good approximation for early actinides but an obviuosly wrong one for late actinides starting from Am and Pu in delta phase, where 5f-electrons are completely or partially localized. 

\begin{figure}[!t]
\begin{center}
\epsfxsize=14cm
\epsfbox{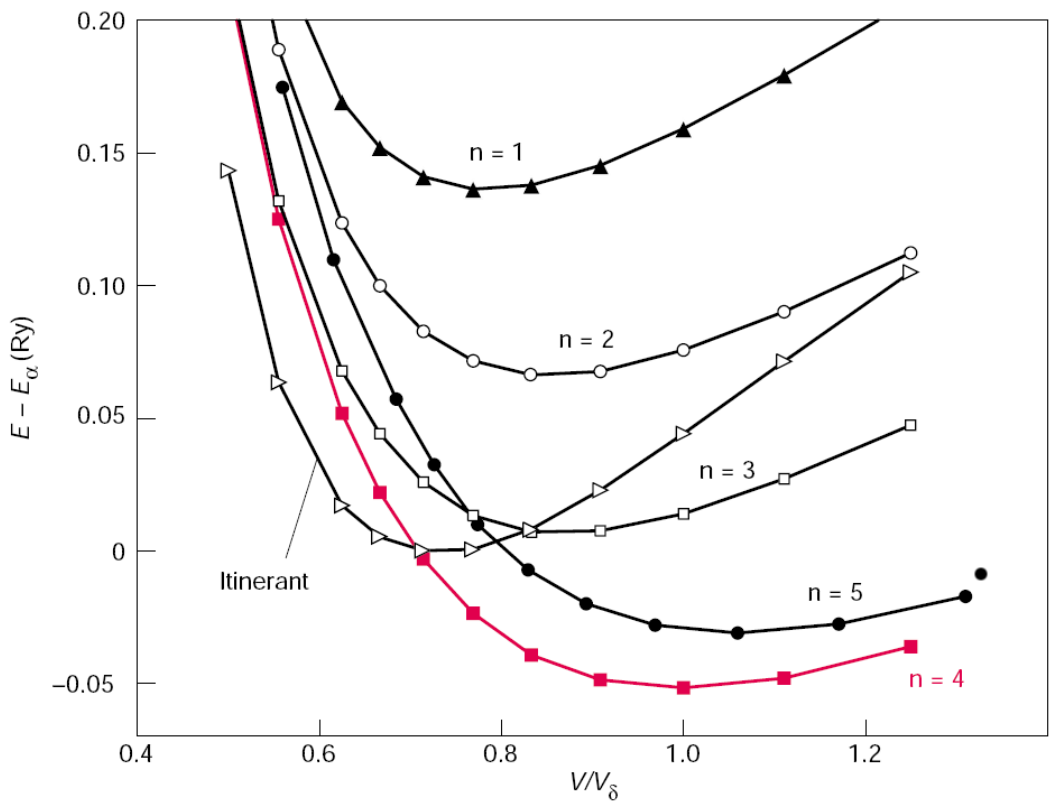}
\end{center}
\caption{Total energy as a function of volume for Pu in delta phase from DFT calculations with various number of 5f-electrons treated as completely localized (``mixed level'' scheme O. Eriksson et al, \cite{eriksson})   }
\label{fig13}
\end{figure}

The term ``partially localized'' means that electrons on every one of 5f-orbitals can demonstrate properties characteristic for localized and itinerant pattern. That can be understood as in different periods of time electrons can either be sitting on specific atomic sites or being spread over the crystal. This effect can be described by Dynamical Mean-Field Theory (DMFT)\cite{DMFT} with time or energy dependent self-energy operator. In so called ``mixed levels'' scheme of O. Eriksson et al \cite{eriksson} ``partial localization'' was imitated by treating some specific 5f-orbitals fully localized (pseudocore) and all others fully itinerant. It was found that the best agreement for equilibrium volume value for Pu in delta phase  was obtained (Fig.\ref{fig13}) with four 5f-electrons considered localized. Similar approach was used in so called Self Interaction Correction (SIC) calculation scheme where calculations gave the best description of Pu in delta phase with localization of three 5f-electrons\cite{Petit01}.  

\begin{figure}[!t]
\begin{center}
\epsfxsize=14cm
\epsfbox{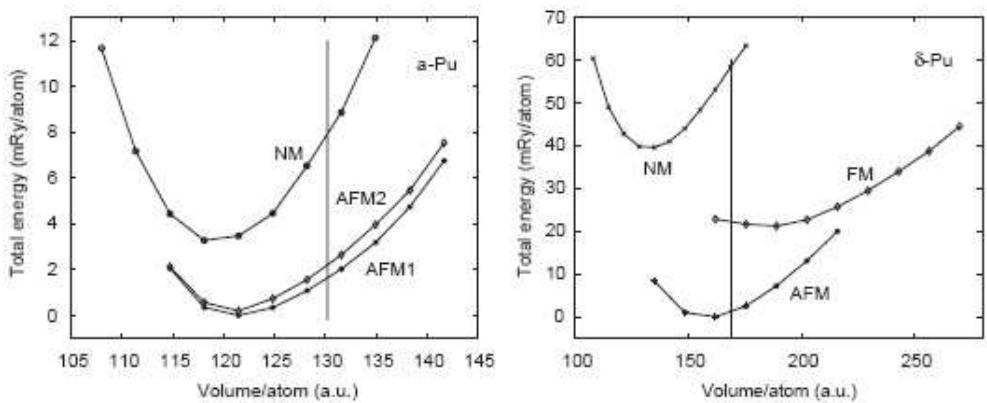}
\end{center}
\caption{Total energy as a function of volume for alpha and delta plutonium from spin-polarized DFT calculations of A.Kutepov et al \cite{Kutepov04}.}
\label{fig3}
\end{figure}

\begin{figure}[!t]
\begin{center}
\epsfxsize=14cm
\epsfbox{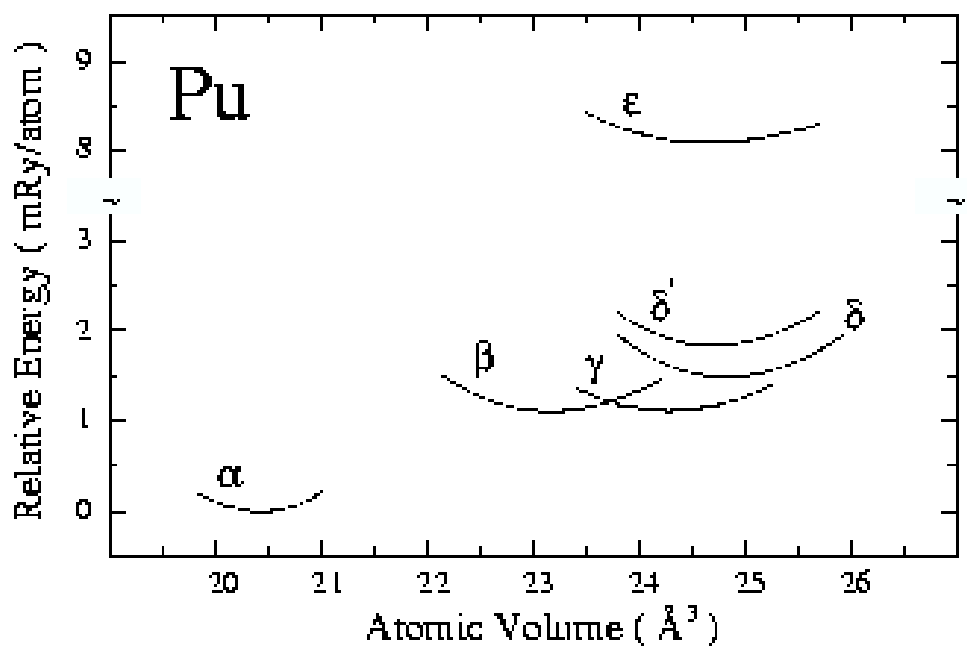}
\end{center}
\caption{Total energy as a function of volume for various crystal structure phases of plutonium from spin-polarized DFT calculations of P.Soderlind et al \cite{soderlind04}.}
\label{fig4}
\end{figure}

\begin{figure}[!t]
\begin{center}
\epsfxsize=14cm
\epsfbox{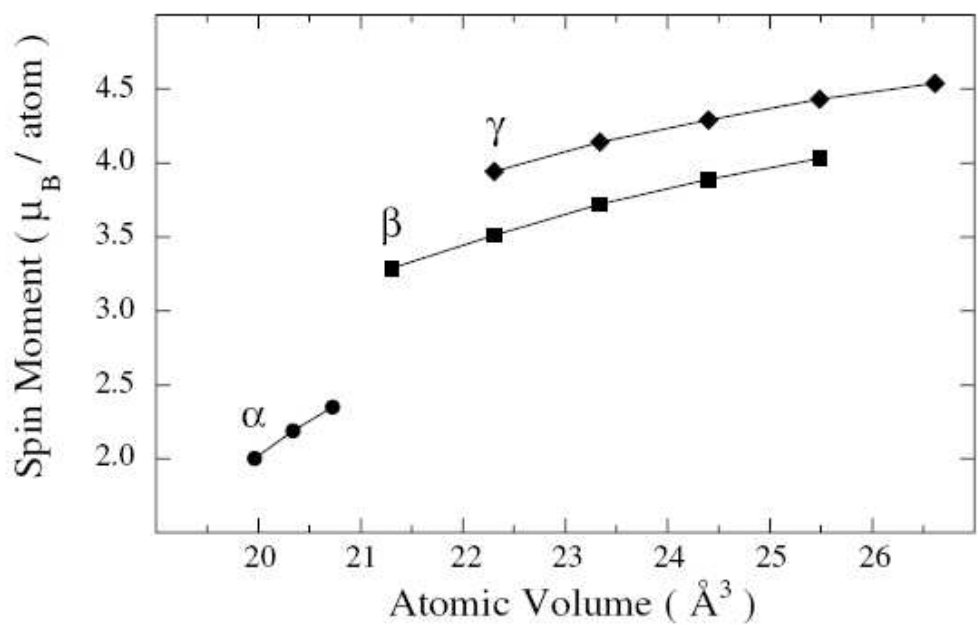}
\end{center}
\caption{Magnetic moment for Pu in various crystal structure phases from spin-polarized LDA calculations (P. Soderlind et al \cite{soderlind04})}
\label{fig12}
\end{figure}

Recently it was found that combination of Generalized Gradient Approximation (GGA)) and spin-polarization taken into account results in drastic improvement of DFT results for equilibrium volume values \cite{Kutepov03,Kutepov04,soderlind04}. While for alpha phase non-magnetic and magnetic solutions  gave total energy as a function of volume curves with close minimum positions, the delta phase calculations show sizable increasing of calculated equilibrium volume value when spin polarization was taken into account (Fig.\ref{fig3}). Not only delta phase but also other crystal structure phase have their calculated volume values in much better agreement with experiment (Fig.\ref{fig4}). The reason for such effect can be understood by observing that phases with large crystal volume  have much larger magnetic moments values (Fig.\ref{fig12}). While for low volume alpha phase spin magnetic  moment values per Pu ion are $\approx 2 \mu_B$, in large volume delta phase the corresponding value is $\approx 4 \mu_B$ \cite{Kutepov03}. Spin polarization results in exchange splitting of 5f-orbitals and hence partial removal of 5f states from the Fermi level. That leads to suppression of 5f contribution to the chemical bonding and increasing of calculated equilibrium volume values. One can say that strong spin polarization  leads to ``partial localization'' of 5f-electrons. This effect is most pronounced in the case of Curium \cite{Cm} where configuration $f^7$ results in full spin polarization with $S=7/2$ and maximum value of exchange splitting for 5f-orbitals with complete removal of 5f states from the Fermi energy. 

The price for this improvement of DFT results is magnetic ground state of Plutonium obtained in calculations in disagreement with experiment. In the recent review of J.C.Lashley et al \cite{Lashley} it was shown that not only there is no experimental evidence for long range magnetic ordering but also the possibility of disordered local moments presence in all Pu phases can be excluded by the results of whole set experimental data including magnetic susceptibility, NMR, specific heat in magnetic field, neutron elastic and inelastic scattering. Recent results \cite{muon} of muon spin relaxation measurements on elemental Pu set an upper limit on ordered moments for alpha-Pu and Ga-stabilized delta-Pu at T=4K of 0.001 $\mu_B$. This disagreement of spin-polarized DFT calculation results with magnetic measurements can not be explained by cancellation of spin magnetic moments by the antiparallel orbital moment \cite{Savrasov00}. The spatial dependence of the orbital and spin
magnetizations is different around the nucleus and if  the total moment is equal to zero the
difference in their spatial extent would still allow a measurable
signal to be seen in neutron scattering \cite{Lashley}. Also the arguments for magnetic moment cancellation in  \cite{Savrasov00}  were based on using formula  $M_{tot}=\mu_B(L-2S)$ that gives zero with $L=5$ and $S=5/2$. However formal total moment value for such case is $J=5/2$. 

\begin{figure}[!t]
\begin{center}
\epsfxsize=14cm
\epsfbox{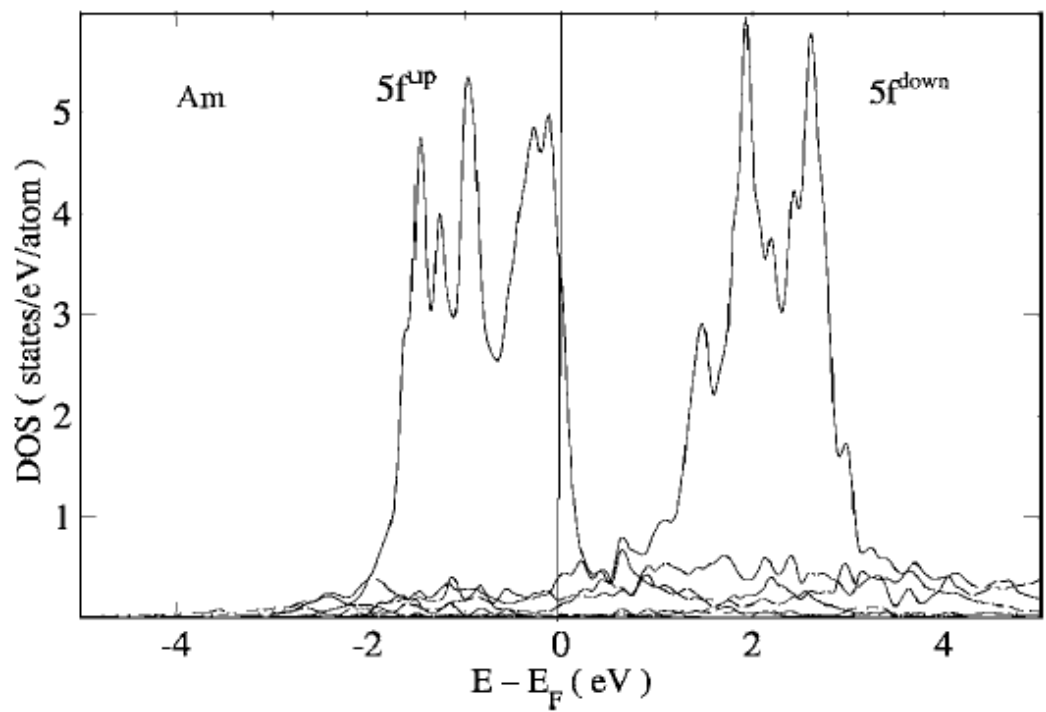}
\end{center}
\caption{5f-electrons partial density of states for Am from spin-polarized LDA calculations (P. Soderlind et al \cite{soderAm})}
\label{fig9}
\end{figure}

\begin{figure}[!t]
\begin{center}
\epsfxsize=14cm
\epsfbox{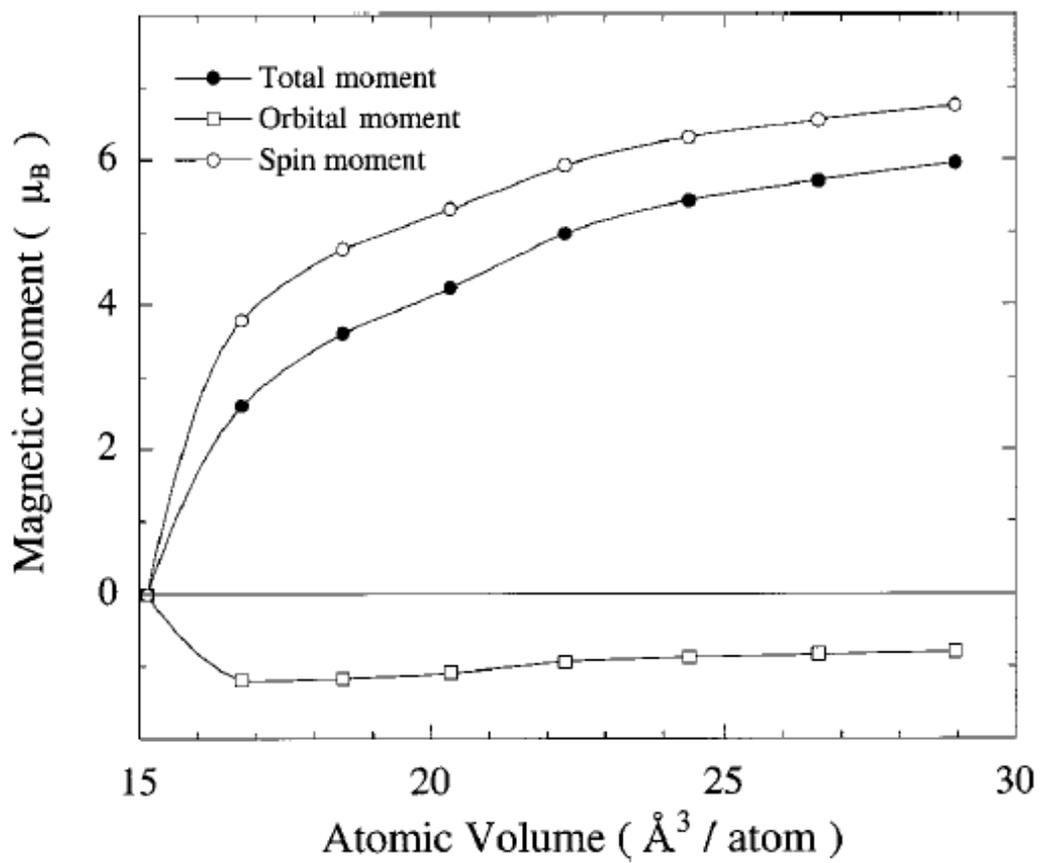}
\end{center}
\caption{Magnetic moment for Am with spin and orbital contributions from spin-polarized LDA calculations (P. Soderlind et al \cite{soderAm})}
\label{fig10}
\end{figure}

\begin{figure}[!t]
\begin{center}
\epsfxsize=14cm
\epsfbox{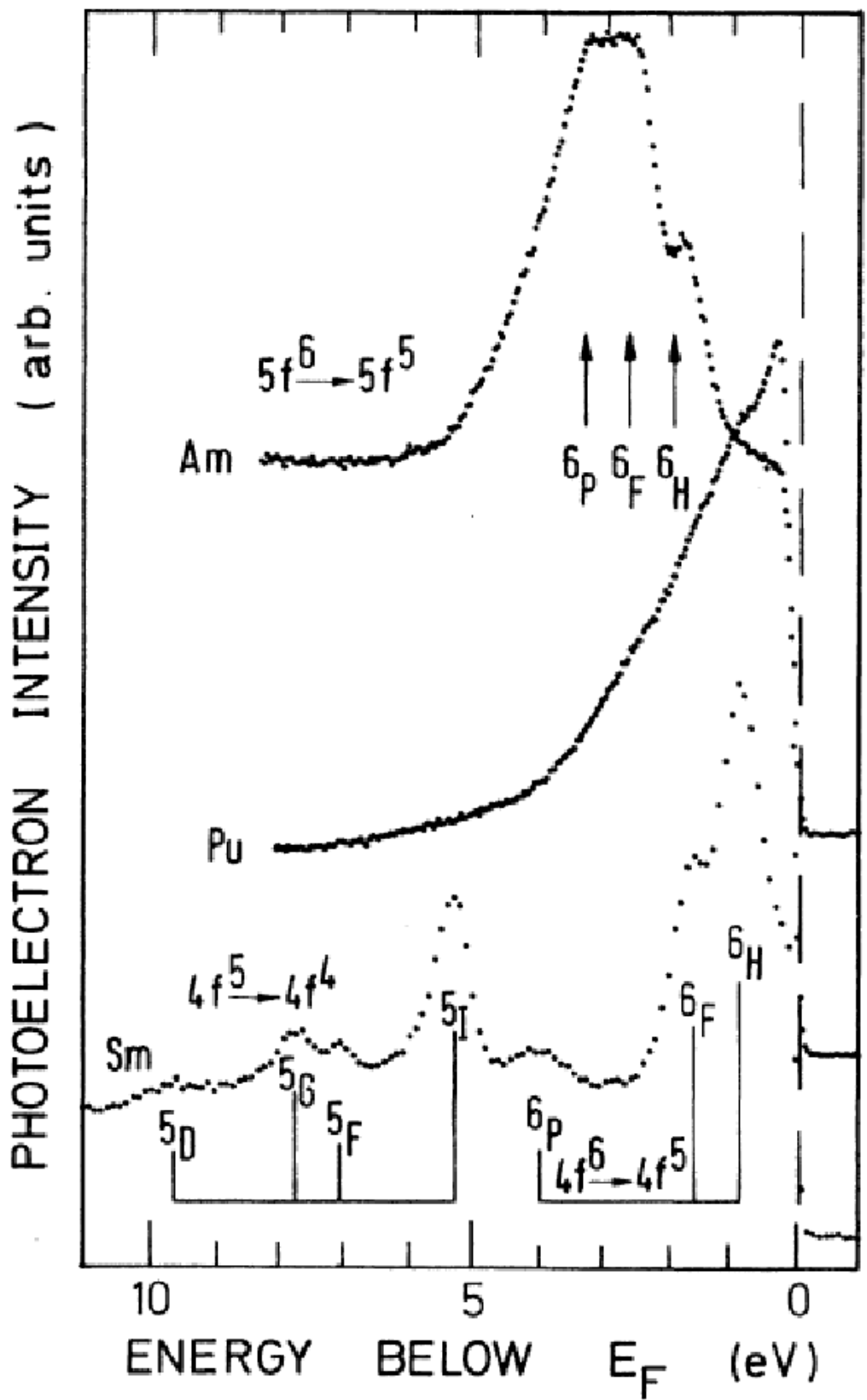}
\end{center}
\caption{Experimental photoemission spectra of Am, Pu and Sm (J.R.Naegele et al \cite{Naegele86})}
\label{fig8}
\end{figure}

In order to understand the origin of spin polarized DFT calculation problems it is instructive to consider much simpler case than Pu with its ``partial localization'' and complicated phase diagram. Americium at ambient pressure has simple hexagonal close-packed crystal structure with localized 5f-electrons in $f^6$ configuration in $jj$ coupling scheme with zero values of spin, orbital and total moment. However DFT calculations \cite{soderAm} gave fully spin polarized solution with huge values of spin and total moments (Fig.(\ref{fig9}-\ref{fig10})). It is not only magnetic properties that DFT gave wrong but also there is a strong disagreement with spectroscopy experiments. In Am photoemission spectra (Fig.\ref{fig8}) occupied 5f band is centered around 3 eV, while in calculated density of states (Fig.(\ref{fig9})) its position is only 1 eV below the Fermi energy. This fact is consequence of full spin-polarization when 5f spin-up sub-band (Fig.(\ref{fig9})) with full capacity 7 electrons is filled with only 6 and Fermi level is inside spin-up sub-band. The ground state obtained in DFT calculations corresponds to the $LS$ coupling scheme with maximum possible spin moment value. In contrast to that in $jj$ coupling scheme the good quantum number is only total moment $J$ and there are $j=5/2$ sub-shell with capacity 6 electrons and $j=7/2$ sub-shell with 8. Then Am with $f^6$ configuration has fully occupied $j=5/2$ sub-shell and empty $j=7/2$ sub-shell with $J=0$. 

The problem of which coupling scheme ($LS$, $jj$ or intermediate) will be realized for particular ion is defined by the competition between Coulomb exchange  interaction preferring maximum spin polarization with strong energy splitting between spin-up and spin-down subbands and spin-orbit coupling leading to energy separation between  $j=5/2$ and $j=7/2$ subsells. While for 3d ions spin-orbit coupling is weak and $LS$ coupling scheme is always valid, for 5f-orbitals of actinides spin-orbit coupling is strong enough to compete with exchange  interaction. In the result for Am with  $f^6$ configuration $jj$ coupling is realized with $S=L=J=0$ and for Cm with $f^7$ configuration $LS$ coupling wins with $S=7/2$. The failure of DFT in describing Am can be due to the overestimation of exchange  interaction strength. In DFT the functional depends on spin-up and spin-down electron densities. This form of the functional suggests  that spin is a good quantum number while for the case of strong spin orbit coupling it can be not true any more.

\section{Coulomb interaction effects in static mean-field approximation: LDA+U calculations}

The problem of ``first principles'' calculations of electronic structure and ground state properties of Plutonium is determined by the question how to describe 5f-electrons localization. The physical origin for localization are correlation effects due to Coulomb interaction between 5f-electrons. In order to include correlation effects into DFT calculations new methods were developed. Two of them are LDA+U\cite{LDA+U} and LDA+DMFT \cite{Anis97,KotliarVollhardt} methods. In the first one  Coulomb interaction is treated in static mean-filed approximation (unrestricted Hartree-Fock) and the second one via Dynamical Mean-Field Theory (DMFT) with energy (or time) dependent complex self-energy operator. 

\begin{figure}[!t]
\begin{center}
\epsfxsize=14cm
\epsfbox{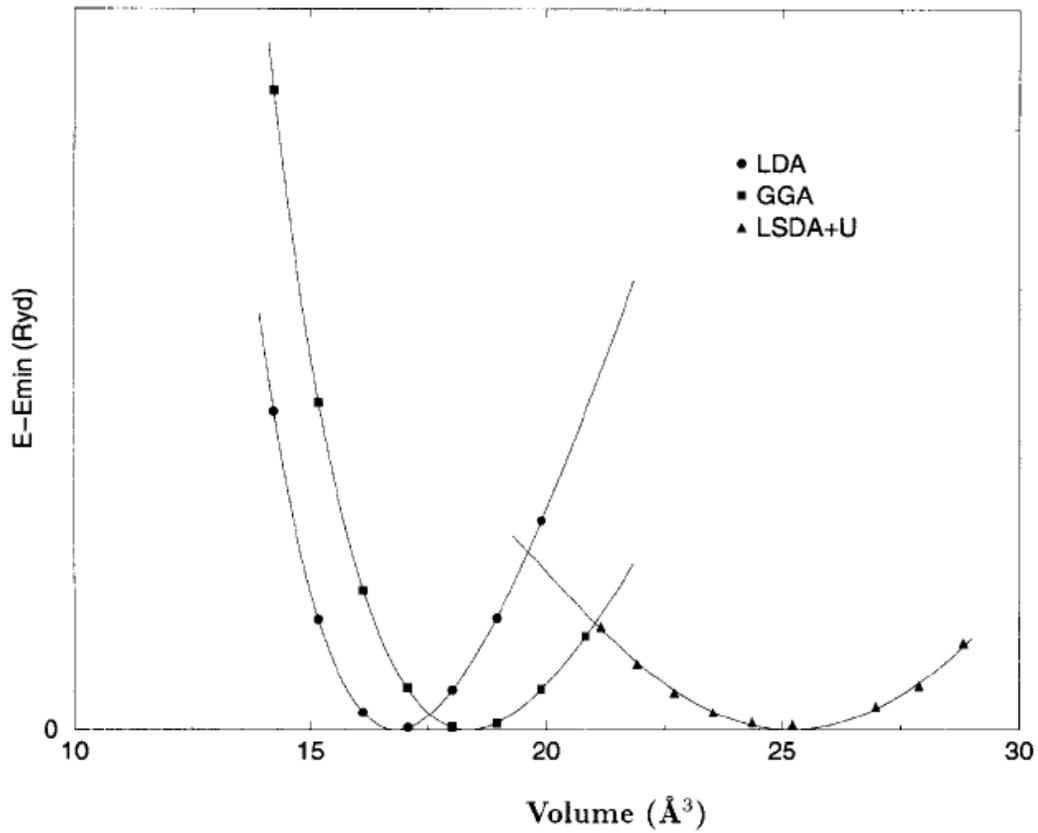}
\end{center}
\caption{Total energy as a function of volume for delta phase of Pu from DFT and LDA+U calculations ( J.Bouchet et al \cite{Bouchet00})}
\label{fig7}
\end{figure}

In LDA+U method potential is orbital dependent with occupied orbitals having lower energy than unoccupied ones.  In the result 5f-orbitals are moved away from the Fermi energy and their contribution to chemical bonding is strongly reduced comparing with paramagnetic DFT results. LDA+U calculations for Pu in delta phase \cite{Savrasov00,Bouchet00} gave significant increasing of equilibrium volume value to a good agreement with experiment (Fig.(\ref{fig7})). However in these works the same problem as in spin-polarized DFT calculations appears: strong spin-polarization with a large values of magnetic moments. 

\begin{figure}[!t]
\begin{center}
\epsfxsize=14cm
\epsfbox{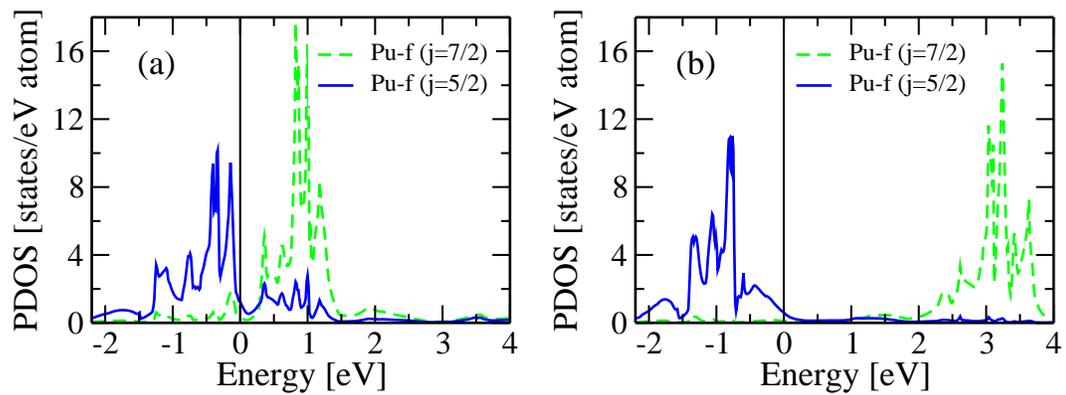}
\end{center}
\caption{Partial 5f densities of states for Pu in delta phase from LDA (a) and LDA+U (b) calculations (A.Shorikov et al \cite{shorikovPu}) }
\label{fig14}
\end{figure}

\begin{figure}[!t]
\begin{center}
\epsfxsize=7cm
\epsfbox{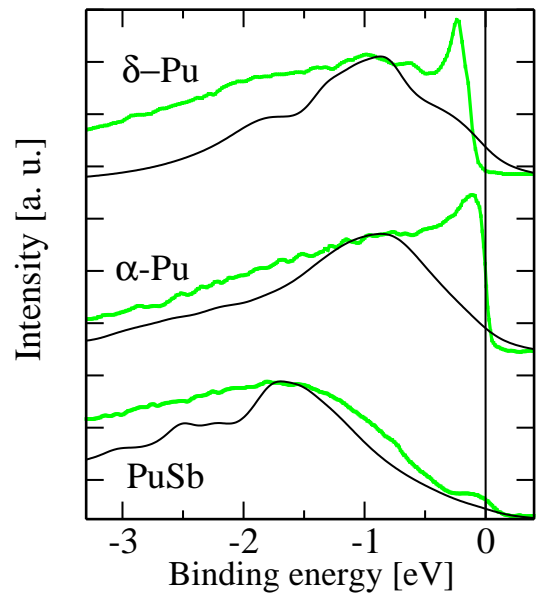}
\end{center}
\caption{Comparison of experimental and calculated (LDA+U) spectra for alpha and delta Pu and PuSb  (A.Shorikov et al \cite{shorikovPu})}
\label{fig15}
\end{figure}

Recently it was found that LDA+U equations can give a nonmagnetic solution\cite{shorikovPu,shick} for Pu with 5f-shell in ground state with $S=L=J=0$ and calculated equilibrium volume for delta Pu in a good agreement with experimental value. The physical origin for such ground state is strong spin-orbit splitting of 5f states that is larger than 5f band width. In pure LDA calculations without LDA+U potential correction (Fig.(\ref{fig14}a)) 5f-orbitals density of states (DOS) consists of two well separated sub-bands being formed predominantly by orbitals with total moment $j=5/2$ for the occupied states and $j=7/2$ for the empty ones. Such 5f-DOS pattern means that nonmagnetic state with filled  $j=5/2$ sub-shell is already nearly preformed in standard LDA description. There is however strong admixture of $j=5/2$ states to the empty band and $j=7/2$ states to the occupied one. When LDA+U potential correction is applied occupied orbitals are shifted down in energy and empty ones up. In the result (Fig.(\ref{fig14}b)) one obtains pure $j=5/2$ and $j=7/2$ sub-bands with increased energy separation between them. The Fermi energy is positioned on the top of the occupied $j=5/2$ sub-band that agrees well with the peak position around 1 eV in photoemission spectra (Fig.(\ref{fig15})).

\begin{figure}[!t]
\begin{center}
\epsfxsize=12cm
\epsfbox{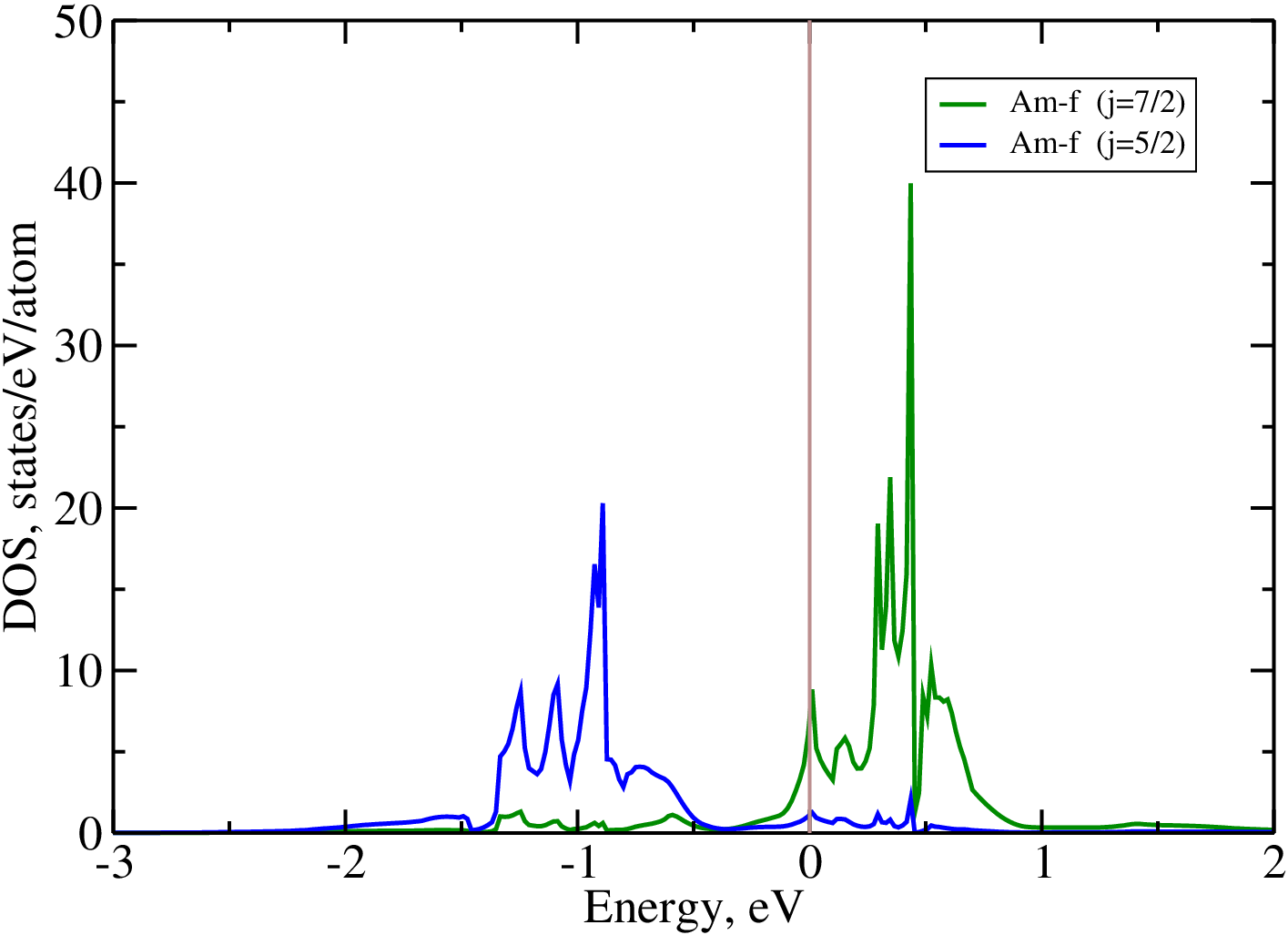}
\end{center}
\caption{5f-electrons partial density of states for Am from LDA calculations (A.Shorikov et al \cite{shorikov}).}
\label{fig6}
\end{figure}

\begin{figure}[!t]
\begin{center}
\epsfxsize=12cm
\epsfbox{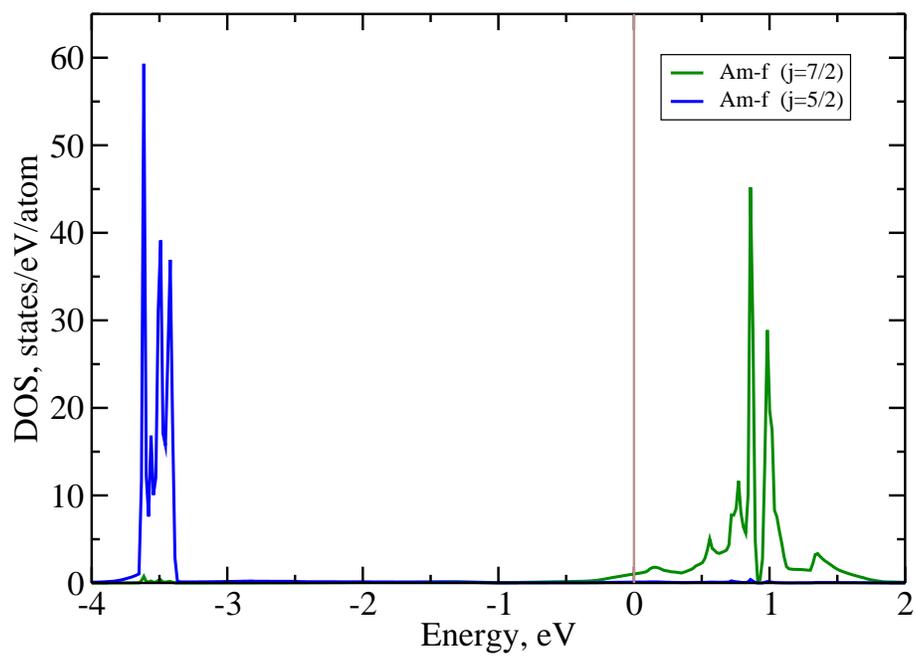}
\end{center}
\caption{5f-electrons partial density of states for Am from LDA+U calculations (A.Shorikov et al \cite{shorikov})}
\label{fig5}
\end{figure}

LDA+U calculations for Am \cite{Am-U,shorikov} gave a nonmagnetic solution with 5f-shell in ground state with $S=L=J=0$. The LDA+U results for Am are in much better  agreement with experiment than spin polarized DFT calculations not only in magnetic  but also in spectral properties. The general 5f-DOS pattern (Fig.(\ref{fig6}-\ref{fig5})) is the same as for Pu (Fig.(\ref{fig14})) with the important difference that position of the Fermi level is not on the top of occupied $j=5/2$ sub-band as it was the case for Pu but at the bottom of empty $j=7/2$ sub-band and smaller 5f band width due to larger volume of Am. This shift of the Fermi level is due to additional valence electron  in Am comparing with Pu which goes into $s-,p-d-$ states not shown on Fig.(\ref{fig6}-\ref{fig5}). In calculated 5f-DOS occupied 5f band is centered around 3 eV in a good agreement with Am photoemission spectra (Fig.\ref{fig8}).

While LDA+U method solved the problem of 5f-electrons localization without developing magnetism that was not observed in experiment there are still important disagreements of LDA+U results with experimental data. First of all it is the absence in the LDA+U calculated spectra of the  sharp peak on the Fermi level observed in experimental photoemission spectra (Fig.(\ref{fig15})). While peak at 1 eV in calculated spectrum for delta Pu agrees well with the corresponding feature in the experimental spectrum, the sharp peak close to the Fermi energy in photoemission spectrum does not find any correspondence in calculations. 

\section{Coulomb interaction effects in dynamic mean-field theory: LDA+DMFT calculations}

This fact can be understood as a manifestation of the ``partially localized'' nature of 5f-electrons in Plutonium. In fully localized case one has occupied lower Hubbard band below the Fermi energy and empty upper Hubbard band above it. If electrons are itinerant then partially filled quasiparticle band crossing the Fermi energy is a correct description. For the intermediate case of ``partial localization'' both features: lower Hubbard band below the Fermi energy and partially filled quasiparticle band would be present in the spectral function, with additional effect of  quasiparticle band narrowing due to correlation effects. That effect can be reproduced in Dynamical Mean-Field Theory (DMFT)\cite{DMFT} with its famous ``three feature structure'' in the spectral function (lower and upper Hubbard bands with quasiparticle peak on the Fermi energy between them). LDA+U method as static mean-field theory can give a good approximation for the Hubbard bands but not the quasiparticle peak. This peak needs dynamical fluctuations to be taken into account for proper description.

Another deficiency of LDA+U solution for Pu is $5f^6$ configuration of the calculated ground state that corresponds to completely filled $j=5/2$ sub-band. From analysis of  absorption spectra of Pu \cite{Laan04} mainly $5f^5$ configuration can be found. The reason for this disagreement could be the following. The Fermi level in LDA+U solution for delta Pu (Fig.(\ref{fig14}b)) is positioned on the top of occupied $j=5/2$ band with the center of the band less than 1 eV below the Fermi energy. That can result in high probability of 5f-electrons excitation from occupied $j=5/2$ band  to the empty $s-,p-d-$ states above the Fermi energy. This is dynamical fluctuations effect and if it will be taken into account the average number of 5f-electrons can be decreased from the static mean-filed solution value.

The disagreement with experiment of LDA+U solution discussed above  leads us to the conclusion that in order do have correct description of 5f states of Pu one should explicitly take into account dynamical fluctuations in calculations. That can be done only in LDA+DMFT method \cite{Anis97,KotliarVollhardt}. The essence of DMFT is mapping of the problem for lattice of atoms with Coulomb interacted electrons on the effective impurity problem for an ion interacting with effective bath (reservoir) characterized by energy (time) dependent hybridization function  that is calculated self-consistently \cite{DMFT}. This effective impurity problem should be solved taking into account full Coulomb correlations between electrons on the ion in reservoir. While being simpler than full lattice problem the impurity problem can be still very expensive computationally, especially for multi orbital case. 

There were developed many methods to solve effective impurity problem, so called ``impurity solvers''(for the recent review on this problem see \cite{RMP06}). One group of them form approximated methods either based on perturbation theories in hybridization strength as Noncrossing Approximation (NCA) or in Coulomb interaction as FLEX. To the same group belongs interpolative approaches like IPT and its extensions. Another group consists of the methods that can be considered formally exact: Quantum Monte Carlo (QMC), Exact Diagonalization (ED) and Numerical Renormalization Group (NRG). The problem with approximated methods is that they give uncontrollable accuracy results. Even if those methods were tested by comparison of their results with exact approaches for some simple cases it is no guaranty that they will work for the problem with a large orbital degeneracy. ED and NRG methods are not practical for orbital degeneracy larger than two. The only method that can in principle give reliable and practically achievable solution is QMC. However for 5f-orbitals with degeneracy equal seven full DMFT-QMC solution can require use of the most powerful computers.

\begin{figure}[!t]
\begin{center}
\epsfxsize=14cm
\epsfbox{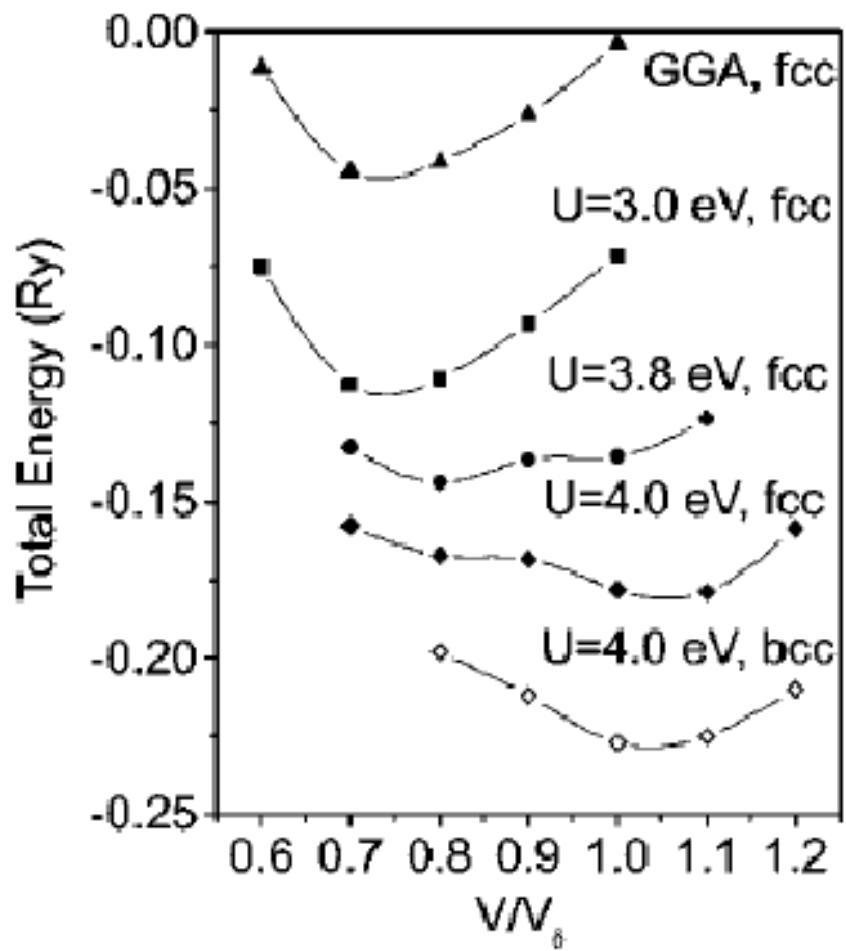}
\end{center}
\caption{Total energy as a function of volume from DMFT calculations (S. Y. Savrasov et al\cite{savrasovDMFT})}
\label{fig17}
\end{figure}

\begin{figure}[!t]
\begin{center}
\epsfxsize=14cm
\epsfbox{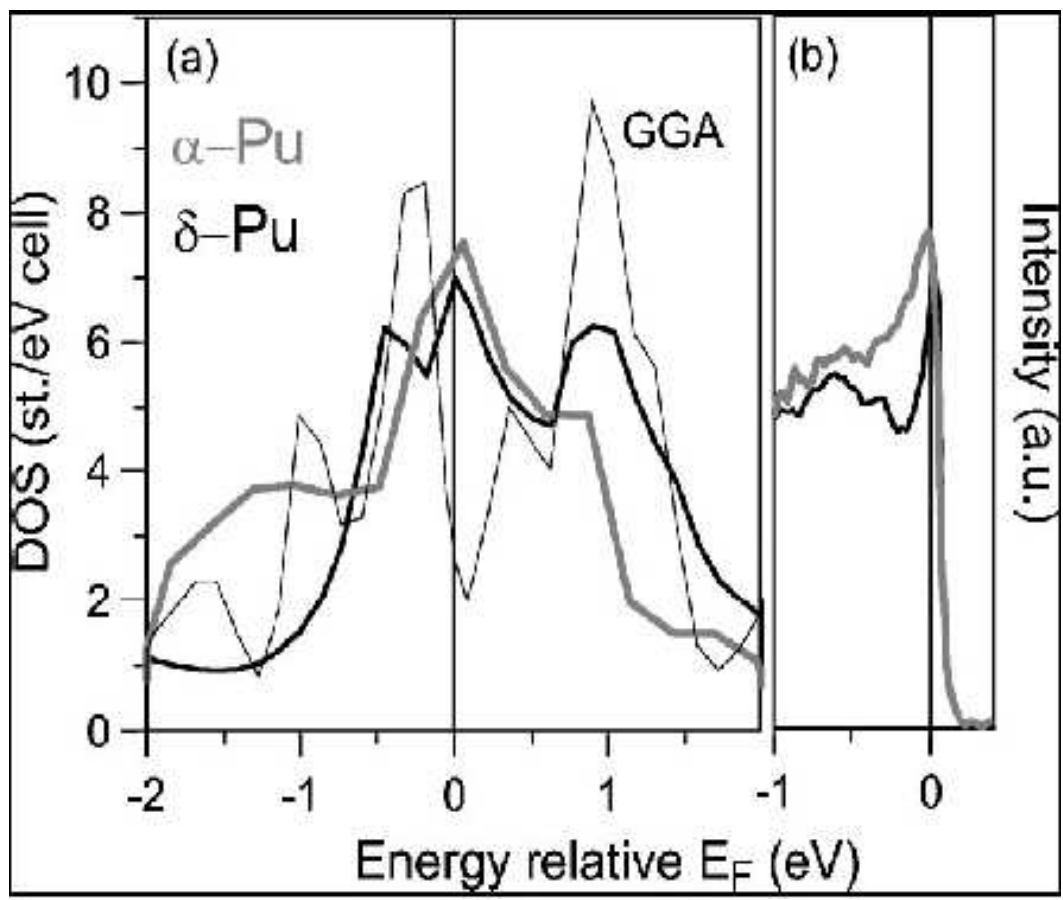}
\end{center}
\caption{Density of states for Pu from DMFT calculations (S. Y. Savrasov et al\cite{savrasovDMFT})
}
\label{fig16}
\end{figure}

LDA+DMFT method\cite{Anis97,KotliarVollhardt} was applied to Pu problem \cite{savrasovDMFT} and encouraging results were obtained including  the possibility of the double minima in energy vs volume curve (Fig.(\ref{fig17})) and the peak on the Fermi level (Fig.(\ref{fig16})). However for ``impurity solver'' the authors used interpolative approach with a simple analytical form for self-energy with parameters being adjusted to obey known various asymptotes. 

\begin{figure}[!t]
\begin{center}
\epsfxsize=14cm
\epsfbox{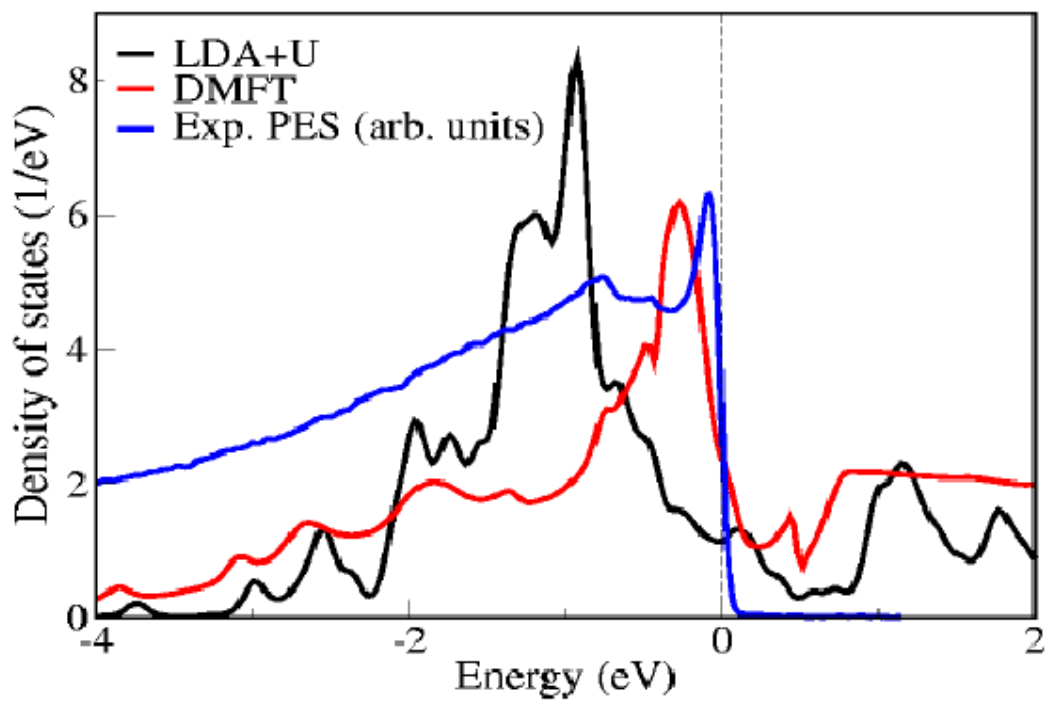}
\end{center}
\caption{Comparison of experimental and calculated (LDA+U, DMFT-FLEX) photoemission spectra for delta Pu (L.V. Pourovskii et al \cite{licht}).}
\label{fig19}
\end{figure}

Another attempt to apply LDA+DMFT method for Pu was done in \cite{licht} where the authors had started from nonmagnetic LDA+U solution and included fluctuation via ``Spin-orbit T-matrix FLEX approach'' based on the perturbation theory in Coulomb interaction strength parameter U. The calculated spectral function (Fig.(\ref{fig19})) shows in comparison with experimental spectrum too strong intensity of the peak on the Fermi energy and suppressed lower Hubbard band.

\begin{figure}[!t]
\begin{center}
\vspace{1cm}
\epsfxsize=14cm
\epsfbox{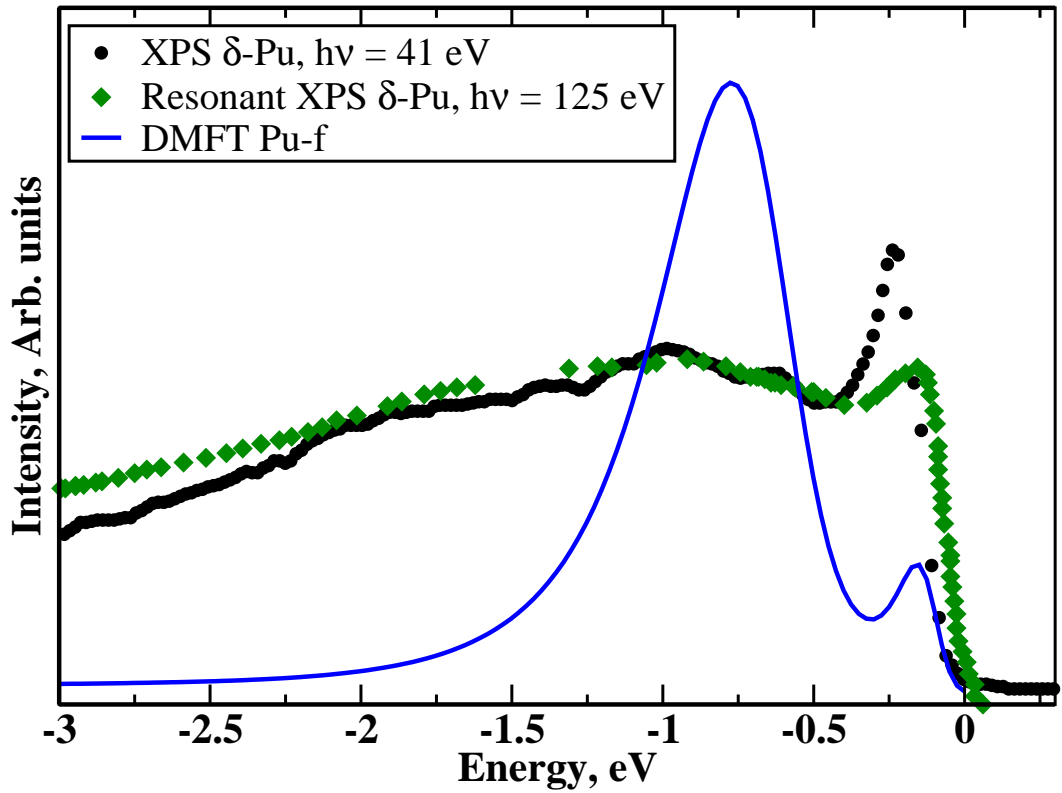}
\end{center}
\caption{Comparison of experimental and calculated (DMFT-QMC) photoemission spectra for delta Pu (J.Kunes et al \cite{kunes}) }
\label{fig18}
\end{figure}

Recently we have done LDA+DMFT calculations for Pu in delta phase with QMC ``impurity solver''\cite{kunes} where only $j=5/2$ 5f-orbitals were treated as fully dynamical while $j=7/2$ 5f-orbitals were described by static mean-filed (Hartree-Fock) approximation. The justification for this was LDA+U solution (Fig.(\ref{fig14}b)) where unoccupied $j=7/2$ band was found to be situated well above the Fermi energy. The calculated spectral function (Fig.(\ref{fig18})) shows correct position of the both features of the experimental photoemission spectra: lower Hubbard band at $\approx$ 1 eV and quasiparticle peak on the Fermi level but the relative intensity of quasiparticle peak was underestimated. The average number of 5f-electrons due to the dynamical fluctuations was significantly decreased from its static mien-filed (LDA+U) value of six electrons to $\approx$5.5 that corresponds to equal weights of $5f^5$ and $5f^6$ configurations in  the ground state.
\section{Conclusion}
In conclusion we have shown that the source of the problem in describing Pu from ``ab-initio'' electronic structure calculations is ``partial localization'' of 5f-electrons. Both limits, itinerant as in DFT and completely localized treating 5f-electrons as pseudocore are not appropriate for Plutonium problem. ``Partial localization'' means that 5f-electrons part of the time spend sitting on particular Pu ion and the rest of the time are spread over the crystal. That corresponds to dynamical (time dependent) fluctuations which can be described by Dynamical Mean-Field Theory (DMFT). Effective impurity problem appearing in DMFT must be solved by Quantum Monet Carlo (QMC) as the only reliable and practically realizable method. However that will require heavy use of  modern multi processor computers.

 This work was partly supported by
RFBI-04-02-16096, 06-02-81017.

\begin{thebibliography}{99}
\bibitem{DFT}
P.~Hohenberg and W.~Kohn,
Phys. Rev. {\bf 136}, B864 (1964);
W.~Kohn and L.~J.~Sham,
{\it ibid.} {\bf 140}, A1133 (1965).
\bibitem{Freeman}
Freeman and Koelling, in {\it The Actinides: Electronic Structure and Related Properties}, edited by A. J. Freeman and J. B. Darby (Academic, New York, 1974) Vol. 1,  51  
\bibitem{DMFT}
A.~Georges, G.~Kotliar, W.~Krauth, and M.~J.~Rozenberg,
Rev. Mod. Phys. {\bf 68}, 13 (1996).
\bibitem{eriksson}
O. Eriksson et al, J. Alloys Compd. 287, 1 (1999); 
\bibitem{Petit01}
L.~Petit, A.~Svane, Z.~Szotek, P.~Strange, H.~Winter, and W.~M.~Temmerman,
J. Phys.: Condens. Matter {\bf 13}, 8697 (2001);
\bibitem{Kutepov03}
A.~L.~Kutepov and S.~G.~Kutepova,
J. Phys.: Condens. Matter {\bf 15}, 2607 (2003).
\bibitem{Kutepov04}
A.~L.~Kutepov and S.~G.~Kutepova,
J. Magn. Magn. Mater. {\bf 272-276}, e329 (2004).
\bibitem{soderlind04}
P. S\"oderlind and B.~Sadigh,
Phys. Rev. Lett. {\bf 92}, 185702 (2004);
\bibitem{Cm}
S.~Heathman, R.G.~Haire, T.Le~Bihan, A.~Lindbaum, M.~Idiri, P.~Normile, S.~Li, R.~Ahuja, 
B.~Johansson, G.H.~Lander, Science {\bf 309},  110-113 (2005)
\bibitem{Lashley}
J.~C. Lashley, A.~C. Lawson, R.~J. McQueeney, and G.~H. Lander, Phys. Rev. B 72, 054416 (2005)
\bibitem{muon}R. H. Heffner et al, 
Phys. Rev. B 73, 094453 (2006)
\bibitem{Savrasov00}
S.~Y.~Savrasov and G.~Kotliar,
Phys. Rev. Lett. {\bf 84}, 3670 (2000).
\bibitem{soderAm}
P. Soderlind et al, Phys. Rev.  61, 8119 (2000);  Phys. Rev.  72, 024109 (2005)
\bibitem{LDA+U}
V.~I.~Anisimov, F.~Aryasetiawan, and A.~I.~Lichtenstein,
J. Phys.: Condens. Matter {\bf 9}, 767 (1997).
\bibitem{Bouchet00}
J.~Bouchet, B.~Siberchicot, F.~Jollet, and A.~Pasturel,
J.~Phys.:~Condens. Matter {\bf 12}, 1723 (2000).
\bibitem{Anis97}V.I.Anisimov, A.I.Poteryaev, M.A.Korotin, A.O.Anokhin, G.Kotliar,
J.Phys.: Condens. Matter {\bf 9} 7359 (1997)
\bibitem{KotliarVollhardt}
G.~Kotliar and D.~Vollhardt,
Phys. Today {\bf 57}, No. 3 (March), 53 (2004).
\bibitem{shorikovPu}
A.O. Shorikov et al, Phys. Rev. B 72, 024458 (2005)
\bibitem{shick}A.B.Shick et al,  Europhys. Lett. 69, 588 (2005)
\bibitem{shorikov}
A.O.Shorikov, V.I.Anisimov, unpublished
\bibitem{Am-U}A.B.Shick et al, Phys. Rev. B 73, 104415 (2006)
\bibitem{Laan04}
G.~van~der~Laan, K.~T.~Moore, J.~G.~Tobin, B.~W.~Chung, M.~A.~Wall, and A.~J.~Schwartz,
Phys. Rev. Lett. {\bf 93}, 097401 (2004);
\bibitem{RMP06} G. Kotliar, S. Y. Savrasov, K. Haule, V. S. Oudovenko, O. Parcollet, and C. A. Marianetti 
Rev. Mod. Phys. 78, 865-951 (2006)
\bibitem{savrasovDMFT} S.~Y.~Savrasov, G.~Kotliar, and E.~Abrahams,
Nature {\bf 410}, 793 (2001);S.~Y.~Savrasov and G.~Kotliar,
Phys. Rev. B {\bf 69}, 245101 (2004);
\bibitem{licht}
L.V. Pourovskii et al, Europhysics Letters 74, 479 (2006).
\bibitem{kunes}
J. Kunes, A.O.Shorikov, V.I.Anisimov, unpublished
\bibitem{LosAlamos}
{\it Challenges in Plutonium Science},
edited by N.~G.~Cooper, Los Alamos Sci. {\bf 26} (LANL, Los Alamos, NM, 2000).
\bibitem{Naegele86}
J.~R.~Naegele, L.~Manes, J.~C.~Spirlet, and W.~M\"uller,
Phys. Rev. Lett. {\bf 52}, 1834 (1986);
\end{thebibliography}
\end {document}